\documentclass[12pt]{article}
\usepackage{amsmath,amssymb}
\usepackage[dvipdfm]{graphicx}
\usepackage{float}

\makeatletter
\renewcommand{\section}{%
 \@startsection{section}{1}{\z@}{-3.5ex \@plus -1ex \@minus -.2ex}{2.3ex \@plus.2ex}%
  {\normalfont\Large\bfseries}}
\makeatother

\makeatletter
\renewcommand{\subsection}{%
 \@startsection{subsection}{1}{\z@}{-3.5ex \@plus -1ex \@minus -.2ex}{2.3ex \@plus.2ex}%
  {\normalfont\large\bfseries}}
\makeatother

\makeatletter
\def\appendix{
\def\theequation{\Alph{section}\arabic{equation}}
\setcounter{equation}{0}
\def\thesection{\Alph{section}}
\@addtoreset{equation}{section}
\setcounter{section}{0}
\def\@seccntformat##1{
\@nameuse{prefix@##1}
\@nameuse{the##1}
\@nameuse{postfix@##1}\quad}
\def\prefix@section{Appendix~}
} 
\makeatother

\makeatletter
\renewenvironment{thebibliography}[1]
{\section*{\refname\@mkboth{\refname}{\refname}}%
   \list{\@biblabel{\@arabic\c@enumiv}}%
        {\settowidth\labelwidth{\@biblabel{#1}}%
         \leftmargin\labelwidth
         \advance\leftmargin\labelsep
	 \setlength\itemsep{+3pt}%
	 \setlength\baselineskip{11pt}%
         \@openbib@code
         \usecounter{enumiv}%
         \let\p@enumiv\@empty
         \renewcommand\theenumiv{\@arabic\c@enumiv}}%
   \sloppy
   \clubpenalty4000
   \@clubpenalty\clubpenalty
   \widowpenalty4000%
   \sfcode`\.\@m}
  {\def\@noitemerr
    {\@latex@warning{Empty `thebibliography' environment}}%
   \endlist}
\makeatother

\def\x{{\boldsymbol x}}

\def\p{{\boldsymbol p}}
\def\E{{\boldsymbol E}}
\def\B{{\boldsymbol B}}
\DeclareMathOperator*{\mysimeq}{\simeq}
\allowdisplaybreaks[4] % permit pagebreak in the gather or the align enviroment

\renewcommand{\thefootnote}{\fnsymbol{footnote}}

\begin{document}

\title{{\Large \textbf{Schwinger mechanism enhanced by the Nielsen--Olesen instability}} }
\author{Naoto Tanji\footnote{\textit{E-mail address}: tanji@post.kek.jp } \ and Kazunori Itakura }
\date{ \normalsize{\textit{High Energy Accelerator Research Organization (KEK), \\
1-1 Oho, Tsukuba, Ibaraki 305-0801, Japan} } } 

\maketitle

\renewcommand{\thefootnote}{$*$\arabic{footnote}}

\begin{abstract}
We discuss gluon production by the Schwinger mechanism in collinear 
color-electric and magnetic fields which may be realized in 
pre-equilibrium stages of ultra-relativistic heavy-ion collisions. 
Fluctuations of non-Abelian gauge fields around a purely color-magnetic 
field contain exponentially growing unstable modes in a longitudinally 
soft momentum region, which is known as the Nielsen--Olesen instability. 
With a color-electric field imposed parallelly to the 
color-magnetic field, we can formulate this instability as the Schwinger 
mechanism. This is because soft unstable 
modes are accelerated by the electric fields to escape from the instability
condition. Effects of instability remain
in the transverse spectrum of particle modes, leading to an anomalously 
intense Schwinger particle production. 
\end{abstract}

\vspace{-380pt}
\begin{flushright}
KEK-TH-1510
\end{flushright}
\vspace{330pt}

%%%%%%%%%%%%%%%%%%%%%%%%%%%%%%%%%%%%%%%%%%%%%%%%%%%%%%%%%%%%%%%%%%%%%%%%%%%%%%%%%%%%%%%%%%%%%%%%%%%%%%%%%%%%%%
\section{Introduction}
\label{intro}

Multi-particle production in strong fields is a typical unstable 
phenomenon that 
can be seen in extreme situations such as ultra-relativistic 
heavy-ion collisions, 
high-energy astrophysical objects, and possibly high-intensity laser. 
This phenomenon is in general nonperturbative 
even for weakly interacting systems because the coupling with external 
strong fields compensates the weakness of the interaction. A straightforward 
description of such an instability may be given by the effective 
action of the strong fields. A famous example is the Euler-Heisenberg action, 
which is an effective action of 
strong electromagnetic fields due to electron's one-loop contribution. 
However, such an effective action does not 
directly tell us the information of dynamical processes which 
cause instability. 
To obtain dynamical pictures of instabilities, 
we have to resort to other methods which allow us to 
describe the degrees of freedom relevant for instabilities.

When the produced particles are normalizable and asymptotically stable modes, one can describe 
particle production within the canonical quantization. 
This is indeed the case with the Schwinger mechanism \cite{Schwinger:1951nm} 
where a pair of a charged particle and an antiparticle is created in a strong field. 

On the other hand, 
when a system contains genuine unstable modes which 
are unnormalizable and keep growing even at asymptotic regions, 
one cannot apply the canonical quantization to those modes. 
When the mode amplitude grows in an unnormalizable way, this mode should 
be treated as not a particle but a field. However, if the instability 
lasts only for a finite duration, we have, after the instability, an 
asymptotic region  where stable particle modes can be defined. 
Then we can provide particle interpretation of the modes by the 
canonical quantization, and therefore the instability can be viewed 
as a particle production phenomena. 
In fact, this situation is seen in the parametric resonance, 
where the time-dependent external field induces instability only for a 
finite time interval, 
and particle number is defined in the free regions
\cite{narozhny1974,mostepanenko1974}. 

Notice that one encounters both of the cases in non-Abelian Yang-Mills theories; 
The one-loop effective action of non-Abelian gauge fields under constant 
electric or magnetic background has an imaginary part, which indicates 
the existence of instabilities \cite{Yildiz:1979vv}. 
The instability in a purely electric field 
consists of normalizable and asymptotically stable modes, and thereby can be interpreted as 
particle productions \cite{Nikishov:1969tt,Tanji:2008ku}. 
One can formulate it as a non-Abelian analog of the Schwinger mechanism \cite{Ambjorn:1982bp}. 
In contrast, the instability in a purely magnetic background is caused 
by rapid growth of particular fluctuations and thus 
cannot be formulated as particle production 
unless the magnetic field is turned off in finite time
so that the modes are stabilized.  
This ``Nielsen--Olesen (N-O) instability" \cite{Nielsen:1978rm} 
is characteristic of non-Abelian gauge fields 
because the properties of self-interaction and spin-1 are necessary.

The instabilities in these two limiting cases have been individually 
studied in the context of early time dynamics of ultra-relativistic heavy-ion 
collisions \cite{Gyulassy:1986jq, Gatoff:1987uf, Fujii:2008dd, 
Iwazaki:2008xi} (see also Ref.~\cite{Berges:2007re}). 
However, the strong non-Abelain fields that appear in reality (called 
a `glasma') would consist of longitudinally extended flux tubes in 
which both the electric and magnetic fields are non-vanishing 
and collinear along the beam axis \cite{Lappi:2006fp}. 
It is important to investigate particle production 
in this configuration to understand time evolution of created matter 
towards a thermalized quark-gluon plasma. In this Letter, we study, 
within the canonical quantization approach, the gauge field instability
(gluon production) 
when both electric and magnetic background fields are present.  

%%%%%%%%%%%%%%%%%%%%%%%%%%%%%%%%%%%%%%%%%%%%%%%%%%%%%%%%%%%%%%%%%%%%%%%%%%%%%%%%
\section{Formalism} \label{formalism}

We consider the SU$(N)$ pure Yang-Mills theory,
and decompose the gauge field $A_\mu ^a$ ($a=1,\cdots,N^2-1$) 
into a classical background $\bar{A} _\mu ^a $ 
and a quantum fluctuation $\mathcal{A} _\mu ^a $ around the background:
$A_\mu ^a = \bar{A} _\mu ^a + \mathcal{A} _\mu ^a $. 
We take a covariantly-constant background field \cite{batalin1977vacuum} 
satisfying
\begin{equation} \label{covariant}
\bar{D} _\mu \bar{F} ^{a \mu \nu} = 0\, ,
\end{equation}
where $\bar{D} _\mu$ is a covariant derivative with respect to the background: 
$\bar{D} _\mu \chi ^a = \left( \partial _\mu \delta ^{ac} -gf^{abc} \bar{A} _\mu ^b \right) \chi ^c $, 
and $\bar{F} _{\mu\nu} ^a $ is a classical field strength: 
$\bar{F} _{\mu\nu} ^a = \partial _\mu \bar{A}_\nu ^a -\partial _\nu \bar{A}_\mu ^a -gf^{abc} \bar{A}_\mu ^b \bar{A}_\nu ^c $.
We employ the background covariant gauge   
\cite{Ambjorn:1982bp}: 
\begin{equation}\label{gauge_fixing}
\mathcal{L} 
 = -\frac{1}{4} F_{\mu\nu}^a F^{a \mu\nu}
   -\frac{1}{2\zeta} \left( \bar{D} _\mu \mathcal{A} ^{a \mu} \right) ^2 
-i\left(\bar{D} ^\mu \bar{C} ^a\right) D_\mu C^a ,
\end{equation}
where $\zeta$ is a gauge parameter (to be set as $\zeta=1$)
and $C^a $ ($\bar{C}^a$) is the Faddeev--Popov (anti-)ghost. 
The last covariant derivative $D_\mu$ contains both the 
background $\bar A_\mu^a$ and the fluctuation ${\cal A}_\mu^a$.
In order to be consistent with the one-loop calculation of 
the effective action, we retain in the Lagrangian up to quadratic 
terms with respect to the quantum fields. 
Then, the equations of motion for the quantum fields are linearized 
as follows: 
\begin{gather}
\bar{D} _\nu \bar{D} ^\nu \mathcal{A} ^{a \mu} -2gf^{abc} \bar{F} ^{b \mu \nu} \mathcal{A} _\nu ^c = 0\, , \label{EOM_A} \\
\bar{D} _\mu \bar{D} ^\mu C^a = 0\, ,  \quad  
\bar{D} _\mu \bar{D} ^\mu \bar{C}^a = 0\, . \label{EOM_barc}
\end{gather}

We consider the classical background field which gives spatially 
constant field strength. Then, one can express the field strength as 
\cite{batalin1977vacuum, Gyulassy:1986jq}
\begin{equation} \label{field_strength1}
\bar{F} _{\mu \nu } ^a = \bar{F} _{\mu \nu } n^a ,
\end{equation}
where $\bar{F} _{\mu \nu }$ is an Abelian field strength and $n^a$ is a color vector, normalized as $n^a n^a = 1 $. %, both of them are independent of $x$. 
The field strength \eqref{field_strength1} is given by the gauge potential
\begin{equation}
\bar{A} _\mu ^a = \bar{A} _\mu n^a ,
\end{equation}
where $\bar{A} _\mu $ is an Abelian gauge field giving 
$\bar{F} _{\mu \nu } = 
\partial _\mu \bar{A} _\nu -\partial _\nu \bar{A} _\mu $. 
This is actually a solution to Eq.~\eqref{covariant}. 
The global residual gauge symmetry allows us to rotate the 
color vector $n^a$ into the Cartan subspace of 
SU$(N)$: $U n^a T^a U^\dagger = \tilde{n} ^A H^A$, 
where $U$ is a constant SU$(N)$ matrix, 
$T^a$ are generators of SU$(N)$ 
and $H^A$  $(A=1,2,\cdots ,N-1)$ are elements of the Cartan 
subspace: $H^A \in \{T^a|\, [T^a,T^b]=0 \}$.

We expand the SU$(N)$ space by $H^A$  and $v_\alpha ^a T^a$, 
instead of $T^a$. Here, 
$v_\alpha ^a$ are eigenvectors of $\text{ad}\{ H^A \} ^{bc} $ 
(i.e., $H^A$  in the adjoint representation): 
\begin{equation}
\text{ad}\{ H^A \} ^{bc} v_\alpha ^c = \alpha ^A v_\alpha ^b \, . 
\end{equation}
The corresponding eigenvalues $\alpha ^A$ may be regarded as an  
$(N-1)$-dimensional vector, which is called the root vector. 
According to this Cartan decomposition, we re-organize the 
quantum fields $\mathcal{A} _\mu ^a $, $C^a$ and $\bar{C} ^a$ as
\begin{align}
\left\{ \mathcal{A} _\mu ^a \right\} &\Rightarrow 
 \left\{ a_\mu ^A %:a=1,\cdots ,N -1 
\right\} \oplus 
 \left\{ W_\mu ^\alpha \equiv v_\alpha ^a \mathcal{A} _\mu ^a \right\} , \\
\left\{ C^a \right\} &\Rightarrow 
 \left\{ c^A %:a=1,\cdots ,N-1 
\right\} \oplus 
 \left\{ \eta ^\alpha \equiv v_\alpha ^a C^a \right\} , \\
\left\{ \bar{C} ^a \right\} &\Rightarrow 
 \left\{ \bar{c} ^A %:a=1,\cdots ,N-1 
\right\} \oplus 
 \left\{ \bar{\eta } ^\alpha \equiv v_\alpha ^a \bar{C} ^a \right\} . 
\end{align}
This decomposition simplifies the interaction 
between the fluctuations $\mathcal{A} _\mu ^a $, $C^a$, $\bar{C} ^a$
and the classical field $\bar{A} _\mu ^a$. %: 
The fields in the Cartan subspace, 
$\phi ^A \equiv a_\mu ^A ,c^A , \bar{c} ^A $, 
are free from the background field: 
$\bar{D} _\mu \phi ^A = \partial _\mu \phi ^A $,
and thus we need not consider these fields in the following. 
For the off-diagonal fields, 
$\Phi ^\alpha \equiv W_\mu ^\alpha , \eta ^\alpha ,\bar{\eta} ^\alpha$, 
the covariant derivative reduces to an Abelian form as
\begin{equation}
\bar{D} _\mu \Phi ^\alpha = \left( \partial _\mu +ie_\alpha \bar{A} _\mu \right) \Phi ^\alpha ,
\end{equation}
with $e_\alpha =-\tilde{n} ^A \alpha ^A g$. 
Gauge invariance of $e_\alpha $ has been explicitly verified for SU(3) \cite{Nayak:2005yv}. 
One can discuss quark production in a similar way \cite{Tanji:2010eu}. 

In terms of these new fields, 
the equations of motion \eqref{EOM_A}, \eqref{EOM_barc} are rewritten as
\begin{gather}
\left( \partial _\nu +ie_\alpha \bar{A} _\nu \right) ^2 W_\mu ^\alpha +2ie_\alpha \bar{F} _{\mu \nu} W^{\alpha \nu} = 0\, , \label{EOM_W} \\
\left( \partial _\mu +ie_\alpha\bar{A} _\mu \right) ^2 \eta ^\alpha = 0\, , 
\quad
\left( \partial _\mu +ie_\alpha\bar{A} _\mu \right) ^2 \bar{\eta } ^\alpha = 0\, . \label{EOM_eta}  
\end{gather}
Solutions of these equations may be expanded by normal modes: 
\begin{gather}
W_\mu ^\alpha (x) 
 =\!\!\!\! \sum _{\sigma =\pm ,\text{L,S} } \sum _n 
   \left[ f_n ^{(\sigma )} (x) a_n ^{(\sigma ) \alpha}  
         + f_n ^{(\sigma )*} (x) b_n ^{(\sigma ) \alpha \, \dagger} \right] \epsilon _\mu ^{(\sigma)} , \nonumber \\
\eta ^\alpha (x) 
 = \sum _n \left[ h_n (x) c_n ^\alpha + h_n ^* (x) d_n ^{\alpha\, \dagger} \right] , \label{c_expansion1} \\
\bar{\eta } ^\alpha (x) 
 = \sum _n \left[ h_n (x) \bar{c} _n ^\alpha + h_n ^* (x) \bar{d} _n ^{\alpha \, \dagger} \right] , \label{barc_expansion1}
\end{gather}
where $n$ is quantum number(s) characterizing the mode, such as momentum 
$\p$, 
and $\sigma (=\pm $,L,S) denotes polarization. % states. 
$f_n ^{(\sigma)} (x) \epsilon _\mu ^{(\sigma)} $ and $h_n (x)$ are c-number solutions of the equations of motion 
\eqref{EOM_W} and \eqref{EOM_eta}, respectively. 
$\epsilon _\mu ^{(\sigma)}$ are polarization vectors satisfying the orthonormal condition
$\epsilon _\mu ^{(\sigma) *} \epsilon ^{(\tau) \mu} = -\tilde{\eta} ^{\sigma \tau}$ and
the completeness condition 
$\sum _{\sigma ,\tau } \epsilon _\mu ^{(\sigma )} \tilde{\eta} ^{\sigma \tau} \epsilon _\nu ^{(\tau ) *} = -g_{\mu \nu}$
with the metric $\tilde{\eta} ^{\sigma \tau}$ such that
$\tilde{\eta} ^{++} = \tilde{\eta} ^{--} = \tilde{\eta} ^\text{LS} = \tilde{\eta} ^\text{SL} = 1 $ and other components are vanishing. 
We identify the solutions $f_n ^{(\sigma)} (x) \epsilon _\mu ^{(\sigma)} $ and $h_n (x)$ as positive frequency modes, and
$f_n ^{(\sigma)*} (x) \epsilon _\mu ^{(\sigma)} $ and $h_n ^* (x)$ as negative frequency modes. 
The positive frequency solutions are to satisfy the following normalization conditions:
\begin{gather}
-g^{\mu \nu} \left( f_n ^{(\sigma)} \epsilon _\mu ^{(\sigma)},f_m ^{(\tau)} \epsilon _\nu ^{(\tau)} \right) 
 = \tilde{\eta} ^{\sigma \tau} \delta _{nm} , \label{norm_f} \\
(h_n ,h_m ) = \delta _{nm} , \label{norm_h} 
\end{gather}
where the inner product is defined as
\begin{equation} \label{inner}
\left( \phi _1 ,\phi _2 \right) 
 = i\int \! d^3 x \, \left( \phi _1 ^* \cdot \bar{D} _0 \phi _2 -\bar{D} _0 \phi _1 ^* \cdot \phi _2 \right) . 
\end{equation}
Notice that
the norms of the negative-frequency modes have 
the sign opposite to those of the positive-frequency modes, 
and the positive and negative frequency modes are orthogonal to each other.  
With these orthonormal conditions, the canonical commutation 
relations imposed on new fields $W_\mu^\alpha, \eta^\alpha, \bar \eta^a$
yield the following commutation relations:  
\begin{gather}
\left[ a_n ^{(\sigma ) \alpha} ,a_m ^{(\tau ) \beta \dagger} \right] = 
 \left[ b_n ^{(\sigma ) \alpha} ,b_m ^{(\tau ) \beta \dagger} \right] = \tilde{\eta} ^{\sigma \tau} \delta _{\alpha \beta} \delta _{nm} , \label{com1} \\
\left\{ c_n ^\alpha ,\bar{c} _m ^{\beta \dagger} \right\} = 
 \left\{ d_n ^\alpha ,\bar{d} _m ^{\beta \dagger} \right\} = i\delta _{\alpha \beta} \delta _{nm} . \label{com2} 
\end{gather}
Now we can regard the operators 
$a_n ^{(\sigma ) \alpha} , b_n ^{(\sigma ) \alpha} , c_n ^\alpha , d_n ^\alpha , \bar{c}_n ^\alpha$ and $\bar{d}_n ^\alpha$ 
as annihilation operators for particles or antiparticles.

%%%%%%%%%%%%%%%%%%%%%%%%%%%%%%%%%%%%%%%%%%%%%%%%%%%%%%%%%%%%%%%%%%%%%%%%%%%%%%%%%%%%%%%%%%%%%%%%%%%
\section{Instability and gluon production} \label{instability}

With the canonical framework defined above, we investigate 
the quantum dynamics of fluctuations in 
the collinear color-electric and magnetic background field:
\begin{equation} \label{config} 
\E ^a = (0,0,E) n^a , \hspace{10pt} \B^a = (0,0,B) n^a ,
\end{equation}
which is realized by $\bar{A} ^\mu = ( 0,-By , 0,-Et) $. 
It is convenient to choose the polarization vectors as eigenvectors of the field strength $\bar{F} ^{\mu \nu}$. 
Then, all the equations that the mode functions 
$f_n ^{(\sigma)} (x) $ and $h_n (x)$ follow can be put together into 
a Klein--Gordon-like equation
\begin{equation} \label{KG}
\left[ \left( \partial _\nu +ie_\alpha \bar{A} _\nu \right) ^2 +m^2 \right] 
\Phi (t,\x  ; m^2 ) = 0 ,
\end{equation}
where 
$\Phi = f_n ^{(\sigma)} , h_n $. 
In the following, the subscript $\alpha $ of $e_\alpha$ will be often omitted.  
The ``mass squared" $m^2$ represents the effects of spin-electromagnetic 
field interaction and is given by
\begin{equation} \label{mass_squared}
m^2 = \left\{
\begin{array}{ccc}
  \mp 2e B  & \hspace{10pt} & \text{for gluon modes with } \sigma = \pm \\
%  2e B   &               & \text{for gluon modes with } \sigma = - \\
  2ie E  &               & \text{for gluon modes with } \sigma = \text{L} \\
  -2ie E &               & \text{for gluon modes with } \sigma = \text{S} \\
  0            &               & \text{for ghost modes}.  
 \end{array} \right. 
\end{equation}
Because the $\sigma =\text{L,S}$ modes acquire pure imaginary $m^2$, 
we should suppose that the negative frequency solution for $\sigma=\text{L}$ 
mode is $f_n ^{(\text{S})*}$, and \textit{vice versa}.

%%%%%%%%%%%%%%%%%%%%%%%%%%%%%%%%%%%%%%%%%%%%%%%%%%
Let us first consider the case where we have only a 
color-electric field ($E\neq 0, B=0$).
Ambj{\o}rn and Hughes \cite{Ambjorn:1982bp} have studied 
the Schwinger particle production in the canonical 
quantization approach in the background covariant 
gauge~(\ref{gauge_fixing}).
One can find exact solutions of Eq.~\eqref{KG} and show that 
all the mode functions satisfy the normalization conditions 
\eqref{norm_f} and \eqref{norm_h}. 
Therefore, in the pure electric field the canonical quantization 
can be completed to give particle interpretation of the fields. 
However, under electric fields, selection of the positive and 
negative frequencies is not unique. 
There are infinite numbers of solutions satisfying the 
conditions \eqref{norm_f} and \eqref{norm_h}. 
If there is no electric field, we can choose the positive and 
negative frequencies uniquely as ${\rm e}^{\mp i\omega _p t}$
respecting the translational invariance in time. 
In contrast, under electric fields, the system loses that 
invariance, so that  
we cannot select a specific set of solutions without other criterion. 
To decide proper solutions, asymptotic WKB criterion is 
employed \cite{Nikishov:1969tt,Ambjorn:1982bp,Tanji:2008ku}. 
In this criterion, the positive and negative frequency solutions 
are selected so that they behave at asymptotic region, 
$|t|\to \infty$, as $\Phi (t,\x  ) \sim \exp (iS_{cl} )$ 
with $S_{cl}$ being the classical action of a charged particle 
under the electric field. 
As a direct consequence of particle production, 
positive and negative frequency solutions defined at $t\to -\infty$ and 
those at $t\to +\infty$ are different. 
The former, which are referred to as in-solutions, are
\begin{equation} \label{in}
\Phi _\p  ^\text{in} (t,\x \, ; m^2) \!
=\! \frac{{\rm e}^{-\frac{\pi}{4} a_{\!\perp} }}{(2e E)^\frac{1}{4} } 
 D_{ia_{\!\perp} -\frac{1}{2} } \left( -{\rm e}^{-\frac{\pi}{4} i} \xi \right) 
   \frac{{\rm e}^{i\p  \cdot \x  }}{\sqrt{(2\pi )^3}} , 
\end{equation}
with $a_{\!\perp} = \frac{m_{\!\perp} ^2}{2e E} $, $m_{\!\perp} ^2 =m^2 +p_x ^2 +p_y ^2$ 
and $\xi = \sqrt{\frac{2}{e E}} (p_z +e Et)$. 
$D_\nu (x) $ is a parabolic cylinder function. 
The latter, out-solutions, are 
\begin{equation} \label{out} 
\Phi _\p  ^\text{out} (t,\x \, ; m^2) 
 = \frac{{\rm e}^{-\frac{\pi}{4} a_{\!\perp} }}{(2e E)^\frac{1}{4} } 
 D_{-ia_{\!\perp} -\frac{1}{2} } \left( {\rm e}^{\frac{\pi}{4} i} \xi \right) 
   \frac{{\rm e}^{i\p  \cdot \x  }}{\sqrt{(2\pi )^3}} .
\end{equation}
Both solutions have the structure of plane waves 
distorted by the electric field. 
Associated with these two kinds of mode functions, we get two kinds of particle creation and annihilation operators, 
$a_\p  ^{(\sigma) \text{in}}$ and $a_\p  ^{(\sigma) \text{out}}$, etc.  
Accordingly, two kinds of vacua $|0,\text{in} \rangle$ and $|0,\text{out} \rangle $ are defined. 
The creation and annihilation operators of in and out are related with 
each other via Bogoliubov transformations. 
Then one can relate the in-vacuum and the out-vacuum and 
calculate the vacuum persistence probability $|\langle 0,\text{out} |0,\text{in} \rangle |^2 $. 
It has been shown that contributions from the unphysical modes, 
$\sigma =\text{L,S}$, and those from the ghost modes are 
totally canceled out, and the pair creation rate $w$ defined by 
$|\langle 0,\text{out} |0,\text{in} \rangle |^2 = \exp (-VT w)$ with $V$ and $T$ being space and time volume, respectively, 
is just twice as large as that of a massless charged scalar field \cite{Ambjorn:1982bp}. 
Also, we can calculate the momentum distribution of out-particles condensed in the in-vacuum,
in other words, the momentum distribution of particles which are created during infinite-time imposition of the electric field. 
For the physical modes, $\sigma =\pm$, we find 
\begin{equation}
\langle 0,\text{in} |
a_\p  ^{(\sigma) \alpha \, \text{out} \dagger} 
a_\p  ^{(\sigma) \alpha \, \text{out} } 
 |0,\text{in} \rangle 
\! = \!
\exp \!\left(\! -\frac{\pi p_{\!\perp} ^2}{e_\alpha E} \right) \! 
\frac{V}{(2\pi)^3} ,\!\!\!
\end{equation}
which depends only on the transverse momentum $p_{\!\perp} ^2 =p_x ^2 +p_y ^2$.%
\footnote{As discussed in Ref.~\cite{Tanji:2008ku}, $p_z$-dependence 
of the distribution cannot be obtained by the method of the asymptotic WKB 
criterion.}  
This result is the same as that of a massless charged scalar 
field. In contrast, for the unphysical modes, $\sigma =\text{L,S}$, and the ghost modes, the expectation is vanishing 
because of their unusual commutation relations \eqref{com1} and \eqref{com2}. 

\begin{figure}[t]
 \begin{center}
  \includegraphics[scale=0.4]{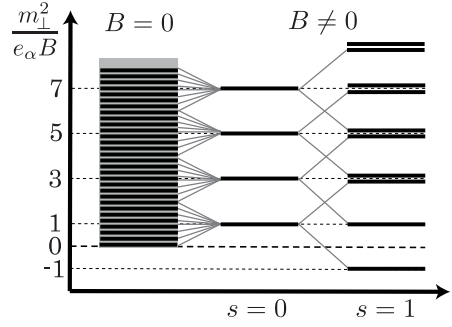} 
 \end{center}
 \vskip -\lastskip \vskip -3pt
\caption{A schematic of the Landau level and the level splitting due to spin-magnetic field interaction
         \label{fig:Landau}}
\end{figure}

%%%%%%%%%%%%%%%%%%%%%%%%%%%%%%%%%%%%%%%%%%%%%%%%%%
In contradistinction to the electric field, the magnetic field has no 
dynamical effect. 
It just discretizes the transverse momenta into the Landau levels as 
\begin{equation} \label{replacement}
p_{\!\perp} ^2 \to (2n+1)e B \hspace{10pt} (n=0,1,2,\cdots ). 
\end{equation}
Under the pure magnetic field ($E=0, B\neq 0$), 
one can find a solution of Eq.~\eqref{KG} as
\begin{equation}
\Phi _\p  (t,\x  ;m^2) 
 = \frac{1}{\sqrt{2\omega _\p  }} {\rm e}^{-i\omega _\p  t} 
   \varphi _n (y) \frac{1}{2\pi} {\rm e}^{ip_x x+ip_z z} ,
\end{equation} 
where 
$\omega _\p  =\sqrt{m^2 +p_z ^2 +(2n+1)e B}$  
and $\varphi _n (y) \propto D_n \left( \sqrt{\frac{2}{e B}} (e By +p_x ) \right) $. 
The spin-magnetic field interaction \eqref{mass_squared} (with $E=0$) shifts 
the Landau levels as shown in Fig.~\ref{fig:Landau}. 
Notice that the transverse mass squared defined as $m_{\!\perp} ^2 = m^2 +(2n+1)e B$ is negative for the lowest Landau level $n=0$ 
with $\sigma =+$ mode, and the one-particle energy reads 
$\omega _{p_z ,n=0} =\sqrt{p_z ^2 -e B} $. 
This is pure imaginary when $|p_z | <\sqrt{eB}$.  
Therefore, the mode functions show exponential growth $\sim \exp \left( \sqrt{e B -p_z ^2} \, t \right)$. 
This is the N-O instability \cite{Nielsen:1978rm}. 
When this instability occurs, the normalization \eqref{norm_f} cannot hold, 
so that we can no longer gain particle interpretation of the field.
If the magnetic field is not constant in time and if there are asymptotically free regions where particle modes can be
defined by the canonical quantization, the N-O instability could be described as particle production phenomena. 
However, as far as the authors know, there has been no explicit demonstration of it.
We will show that we can define asymptotically stable particle modes even in the presence of the constant magnetic field
provided there is a constant electric field parallel to the magnetic field. 

%%%%%%%%%%%%%%%%%%%%%%%%%%%%%%%%%%%%%%%%%%%%%%%%%%
Notice that the N-O instability occurs only at low longitudinal momentum region $|p_z |<\sqrt{eB} $. 
Under the collinear electric and magnetic fields ($E\neq 0, B\neq 0$), particles are accelerated by the electric field
and their momenta increase in time as $p_z =e Et$\,+\,const. 
Therefore, even if a particle lies in the unstable momentum region $|p_z |<\sqrt{eB} $ at some time, 
it escapes from there unfailingly at later time. 
Owing to this mechanism, the exponential growth of the mode function is regularized in the presence of the color-electric field,
and we can carry out the canonical quantization and obtain the particle interpretation for all the modes. 
Indeed, exact solutions of Eq.~\eqref{KG} under the collinear fields 
\begin{gather}
\Phi _\p  ^\text{in} (t,\x ; m^2) \!
 =\! \frac{{\rm e}^{-\frac{\pi}{4} a_n }}{(2e E)^\frac{1}{4} } 
   D_{ia_n -\frac{1}{2} }\! \left( -{\rm e}^{-\frac{\pi}{4} i} \xi \right) \!\varphi _n (y)
   \frac{{\rm e}^{ip_x x +ip_z z }}{2\pi } , \\
\Phi _\p  ^\text{out} (t,\x  ; m^2) \!
 =\! \frac{{\rm e}^{-\frac{\pi}{4} a_n }}{(2e E)^\frac{1}{4} } 
   D_{-ia_n -\frac{1}{2} }\! \left( {\rm e}^{\frac{\pi}{4} i} \xi \right)\! \varphi _n (y)
   \frac{{\rm e}^{ip_x x +ip_z z }}{2\pi } , 
\end{gather}
with $a_n = \left( n+\frac{1}{2} \right) \frac{B}{E}$ 
do not diverge at $t\to +\infty$ and satisfy the normalization conditions \eqref{norm_f} and \eqref{norm_h}. 
This is true even for the lowest Landau mode with $n=0$ and $\sigma =+$.
Since the magnetic field has no dynamic effect, the selection of the in- and out- mode functions is not affected by it. 
Therefore, the Bogoliubov relations are almost the same as those in the pure electric field. 
The differences are the emergence of the Landau levels \eqref{replacement} 
and the level splitting between the two transverse modes %, $\sigma =\pm$ 
(see Eq.~\eqref{mass_squared}). 
Because $m^2$ for the unphysical modes are independent of $B$, 
cancellation between those modes again holds 
even in the presence of the magnetic field. 
Actually, the pair creation rate has contributions only from the two 
physical transverse modes: %($\sigma =\pm $):
\begin{equation} \label{rate}
w = \frac{N}{2} \frac{g^2 EB}{(2\pi)^2} \sum_{\sigma =\pm } \sum _{n=0} ^\infty 
    \ln \left[ 1+{\rm e}^{-\pi (2n+1-2\sigma ) \frac{B}{E} } \right] . 
\end{equation}
Furthermore, the in-vacuum expectations of out-number operators for the 
unphysical modes are all vanishing
also in the presence of the magnetic field. 
Those for the physical transverse modes are
\begin{equation}
\langle 0,\text{in} |a_{p_z ,n} ^{(\pm ) \alpha \, \text{out} \dagger} a_{p_z ,n} ^{(\pm ) \alpha \, \text{out} } |0,\text{in} \rangle 
 =  {\rm e}^{-\pi (2n+1 \mp 2) \frac{B}{E} }
 \frac{V}{(2\pi )^3} . \label{number+}
\end{equation}
In Eqs.~\eqref{rate} and \eqref{number+}, 
the lowest level ($\sigma =+$ and $n=0$) is worthy of great remark. 
In this mode, the effective mass square is negative, $m_{\!\perp} ^2 = -eB$, 
so that the index of the exponential factor is positive: %$\exp ( +\pi B/E )$
\begin{equation} \label{result}
\langle 0,\text{in} |a_{p_z ,n=0} ^{(+) \alpha \, \text{out} \dagger} a_{p_z ,n=0} ^{(+) \alpha \, \text{out} } |0,\text{in} \rangle 
 = {\rm e}^{+\pi \frac{B}{E} } \frac{V}{(2\pi )^3} . 
\end{equation}
This means that an anomalously large number of gluons are produced in the lowest mode 
when compared with the usual Schwinger mechanism of QED or scalar QED. 
This is because the field fluctuations amplified by the N-O instability
are converted to real particles by the electric field. 

The collinear color electromagnetic field \eqref{config} 
may be realized in flux tubes created at the initial stage 
of heavy-ion collisions \cite{Lappi:2006fp}.
The strongly enhanced particle production causes rapid decay of 
those coherent fields into particle degrees of freedom, 
which would be followed by the formation of a quark-gluon plasma. 

%%%%%%%%%%%%%%%%%%%%%%%%%%%%%%%%%%%%%%%%%%%%%%%%%%%%%%%%%%%%%%%%%%%%%%%%%%%%%%%%
\section{Discussion} \label{discussion}

Our analysis has relied on the linear approximation. 
If the field amplification by the N-O instability continues for a long time, the linear approximation would get invalid. 
A typical time scale of the N-O instability is given by the inverse 
of the growth rate $t_{\rm NO} \sim 1/\sqrt{eB}$.
Therefore, the time scale when the non-linear corrections become 
important would be $t_{\rm nl} \gtrsim 1/\sqrt{eB}$. 
Actually, it has been demonstrated by numerical calculations that the 
exponential growth of the gauge fluctuations stops at some time 
(later than $t_{\rm NO}$) due to non-linear corrections \cite{Berges:2011sb}. 
Meanwhile, the time scale of the stabilization process by the electric field 
$t_{\rm ele}$ 
is given from the condition that a particle is accelerated away from the instability condition $|p_z |<\sqrt{eB} $,
so that $t_{\rm ele} \sim \sqrt{eB}/eE $. 
If the electric and magnetic fields are the same order of magnitude, which is a situation expected in a glasma, 
$t_{\rm ele}$ is the same order as $t_{\rm NO}$, 
so that the electric field plays a dominant role to stabilize the N-O 
instability. 

To understand more clearly 
our result that the N-O instability is stabilized by the electric field and induces the strong particle production, 
the following simple toy model which describes particle production by an instability may be useful. 
Let us suppose that, without specification of background fields, 
a real scalar field suffers an instability during a finite period, $0<t<T$. 
The mode functions of the field have the following properties:
(i) for $t<0$ the modes are stable and have the plane wave solutions; 
$\frac{1}{\sqrt{2\omega }} {\rm e}^{\mp i\omega t}$ $(\omega \geq 0)$,
(ii) for $0<t<T$ the modes are unstable and show exponential growth or 
decrease; ${\rm e}^{\pm \gamma t}$ $(\gamma >0)$, and
(iii) for $T<t$ the modes are again stable; $\frac{1}{\sqrt{2\Omega }} 
{\rm e}^{\mp i\Omega t}$ $(\Omega \geq 0)$. 
The canonical quantization can be performed at $t<0$ and at $t>T$, 
and correspondingly two kinds of particle definitions of `in' and `out' are obtained. 
One can find the relation between these two kinds of particle definitions by smoothly connecting the mode solutions 
in the three time regimes. 
If the solution $\frac{1}{\sqrt{2\omega }} {\rm e}^{-i\omega t}$ for $t<0$ is smoothly connected to the linear combination of
the solutions for $0<t<T$, and is subsequently connected to those
for $t>T$ as $\frac{1}{\sqrt{2\Omega }} \left( \alpha {\rm e}^{-i\Omega t} +\beta ^* {\rm e}^{i\Omega t} \right)$, 
then the particle annihilation (creation) operators of in-particle, $a^{\text{in} (\dagger)}$, 
is related with those of out-particle, $a^{\text{out} (\dagger)}$, by the Bogoliubov transformation: 
$a^{\text{out} } = \alpha a^\text{in} +\beta a^{\text{in} \, \dagger}$.
One can explicitly confirm the condition $|\alpha |^2 -|\beta |^2 =1$,
which guarantees the unitarity, holds irrespective of the values of $\omega$, $\Omega$, $\gamma$ and $T$. 
The number of particles created by the instability is
\begin{equation} \label{toy}
\langle 0,\text{in} |a^{\text{out} \dagger} a^\text{out} |0,\text{in} \rangle
 = |\beta |^2 
 \mysimeq _{T\gg 1/\gamma} 
  \ \frac{\Omega }{16\omega } \left( 1+\frac{\omega ^2}{\gamma ^2} \right) \left( 1+\frac{\gamma ^2}{\Omega ^2} \right) {\rm e}^{2\gamma T} . 
\end{equation}
Thus, the number of created particles increases exponentially with 
respect to the time interval of the unstable region, $T$. 
In this simple demonstration, the existence of the stable modes in asymptotic regions has been crucial to formulate the particle production 
in a canonical way. 
If we compare this simple model with our original problem of the gluon 
production, the instability parameter $\gamma $ corresponds to
$1/t_{\rm NO} \sim \sqrt{eB}$. 
Though there is no explicitly stable region in temporary direction under the constant magnetic field, 
the electric field provides us with the stable regions in the momentum space, \textit{i.e.} $|p_z |>\sqrt{eB}$. 
Hence, $T$ should be replaced by $t_{\rm ele} \sim \sqrt{eB}/eE $. 
By substituting these parameters, 
we find the exponential factor ${\rm e}^{2\gamma T} $ in Eq.~\eqref{toy} roughly reproduces the exact result \eqref{result}.

%%%%%%%%%%%%%%%%%%%%%%%%%%%%%%%%%%%%%%%%%%%%%%%%%%%%%%%%%%%%%%%%%%%%%%%%%%%%%%%%%%%%%%%%%
\section*{Acknowledgments}
N.~T. is supported by the Japan Society for the Promotion of Science for Young Scientists.

%%%%%%%%%%%%%%%%%%%%%%%%%%%%%%%%%%%%%%%%%%%%%%%%%%%%%%%%%%%%%%%%%%%%%%%%%%%%%%%%%%%%%%%%%

\end{document}